\documentclass[reprint,amsmath,amssymb,prl,floatfix,twocolumn]{revtex4}

\pdfoutput=1
\usepackage{graphicx}
\usepackage{dcolumn}
\usepackage{bm}
\usepackage{times}
\usepackage[usenames,dvipsnames]{color}
\usepackage{colordvi,epsf}

\begin{document}


\title{Engineering skyrmions in transition-metal multilayers for spintronics}
\author{B.~Dup\'e${}^{1,2}$, G.~Bihlmayer${}^2$, S.~Bl\"ugel${}^2$, and S.~Heinze${}^{1}$}
\affiliation{${}^1$Institute of Theoretical Physics and Astrophysics, University of Kiel,
24098 Kiel, Germany}
\affiliation{${}^2$Peter Gr\"{u}nberg Institut (PGI-1)   \&
Institute for Advanced Simulation (IAS-1), Forschungszentrum J\"{u}lich and JARA,
52425 J\"{u}lich, Germany}

\date{\today}

\begin{abstract}
Magnetic skyrmions are localized, topologically protected spin-structures that
have been proposed for storing or processing information due to their  intriguing dynamical and transport properties.
Important in terms of applications is the recent discovery of interface stabilized skyrmions 
as evidenced in ultra-thin transition-metal films.
However, so far only skyrmions at interfaces with a single atomic layer of a magnetic material were reported,
 which greatly limits their potential for application in devices. 
Here, we predict the emergence of skyrmions in $[4d/\mathrm{Fe}_2/5d]_n$ multilayers, i.e. structures
composed of Fe biatomic layers sandwiched between 
$4d$- and $5d$-transition-metal layers. In these composite structures, the exchange and the Dzyaloshinskii-Moriya interactions
which control skyrmion formation can be tuned separately by the two interfaces. This allows engineering 
skyrmions as shown based on density functional theory and spin dynamics simulations.

\end{abstract}

\pacs{73.20.-r 
      71.15.Mb 
      }
      
\keywords{skyrmions, topology, magnetic multilayers}
\maketitle

Topologically protected spin-structures, in particular magnetic skyrmions \cite{Bogdanov1989,Bogdanov1994,Roessler:06.1}, 
have recently received much attention due to their promising applications 
in spintronics~\cite{Fert:13.1,Kiselev2011,Sampaio:13.1}. 
Among the different mechanisms that can induce a topologically non-trivial 
magnetic structure, the relativistic Dzyaloshinskii-Moriya interaction (DMI) \cite{Dzyaloshinskii,Moriya}
arising in non-centrosymmetric systems is maybe the most prominent one~\cite{Nagaosa:13.1}.
Magnetic skyrmions have been first revealed in a few non-centrosymmetric bulk magnets,
such as the chiral MnSi~\cite{Muehlbauer:09.1,Pappas2009}
or FeCoSi~\cite{FeCoSi1}. In two-dimensional (2D) systems the DMI arises always due to the broken inversion
symmetry at the interface between a magnetic film and a substrate~\cite{Crepieux1998,Bogdanov:2001aa,Bode2007}.
E.g.\ an Fe monolayer on an Ir(111) substrate exhibits strong DMI and on 
this system a nanoscale skyrmion lattice was first observed by scanning-tunneling 
microscopy (STM)~\cite{Heinze:11.1}. This skyrmion lattice is stabilized by the 
short-range four-spin interaction and slight modifications of the film composition,
e.g.\ adding an atomic Pd overlayer, can push this system into a regime where 
individual skyrmions can exist which can be manipulated by spin-polarized currents
in STM~\cite{Romming:13.1}. 

The manipulation of ultrathin films such as Fe/Ir(111) is a versatile 
route to tailor properties such as the size or magnetic field needed to induce a 
magnetic skyrmion~\cite{Dupe:14.1,Simon:14.1}. For practical applications, however, there
are several difficulties that have to be overcome: (i) Manipulation of skyrmions 
by electric currents~\cite{Sampaio:13.1,Jonietz:10.1,Yu:12.1,Everschor2012,Iwasaki:13.1,Iwasaki:13.2} 
and measurement of the topological Hall effect \cite{Lee2009,Neubauer2009,Franz2014}
requires that a substantial fraction of the current
runs through the magnetic structure. In a single Fe layer on a metallic substrate
this is difficult to realize. (ii) Creating large area single Fe layers on a substrate
requires specialized preparation methods that hamper large-scale production of
these structures. (iii) Temperature stability and optimization of the size of these magnetic structures 
requires a tuning of different parameters such as exchange-interactions, the DMI, and
the magnetic anisotropy. 

Here, we propose multilayers composed of Fe bilayers epitaxially sandwiched between
$4d$- and $5d$-transition-metal layers as promising systems towards solving
these issues.  The repetition of layered structures is a powerful method in 
spintronics to engineer materials properties~\cite{Fert2015,Chen2015}:
Besides increasing the magnetic material in $[4d/\mathrm{Fe}/5d]_n$ multilayers, such structures
have the key advantage that the essential magnetic interactions, the 
exchange and the DMI, can be controlled by two separate interfaces.
Thereby, the requirement of pseudomorphic growth is limited to the interface
which induces the DMI while intermixing at the other interface is not crucial. 
Based on density functional theory (DFT), we show a route to engineer the 
skyrmion systems by varying the interface composition.
We focus on an Fe bilayer sandwiched between Ir and Rh/Pd layers,
however, other combinations of $4d$ and $5d$ transition-metals are also feasible.
Using spin dynamics simulations
we find the emergence of skyrmions in the Fe bilayers and study their properties.

In order to obtain the magnetic interactions in transition-metal (TM) multilayers we
have performed DFT calculations for various noncollinear spin structures including
spin-orbit coupling~\cite{Kurz-SS,Heide-DMI}. By mapping the total energies of the
different magnetic configurations to an extended 
Heisenberg model we can extract parameters for the interactions. We consider homogeneous
flat spin spirals with a period $\lambda$ in which the magnetization rotates along a given
crystallographic direction that lies in the plane of the interface
by an angle of $\phi=(2\pi/\lambda) d$ from atom to atom where
$d$ is the spacing between adjacent atoms (see methods).
This approach has been successfully used in the past to study complex spin textures 
in a wide range of systems such as biatomic 
chains at surfaces~\cite{PhysRevLett.108.197204}, ultra-thin film skyrmion
systems~\cite{Heinze:11.1,Dupe:14.1} and  magnetic mono-~\cite{Bode2007,Ferriani-MnW,Zimmermann2014} 
and bilayers~\cite{PhysRevLett.108.087205}. 

A characteristic quantity of the exchange interaction in the considered TM multilayers
is the effective nearest-neighbor exchange constant $J_{\rm eff}$. It is a measure
of how quickly the energy rises for long-wave length spin spirals with respect to the 
ferromagnetic state. $J_{\rm eff}$ was obtained by fitting the dispersion curve $E(\mathbf{q})$ 
of spin spirals characterized by the wave vector $\mathbf{q}$ in the vicinity 
of the ferromagnetic (FM) state ($|\mathbf{q}| = 0$) with a parabola $J_{\rm eff} \frac{3a^2}{2}|q|^2$
where $a$ is the nearest-neighbor distance within the layers
(for energy dispersions see supplementary material S2). 
Therefore, a negative value of $J_{\rm eff}$ indicates a spin spiral ground state driven by 
exchange interaction. As a second characteristic quantity, we include the energy difference
between a spin spiral state with a wave length $\lambda=2 \pi/|q|=2.7$~nm  and
the ferromagnetic state, denoted by $E_{\rm SS}$. 

\begin{figure}[thp]
\centering
\includegraphics[width=0.46\textwidth]{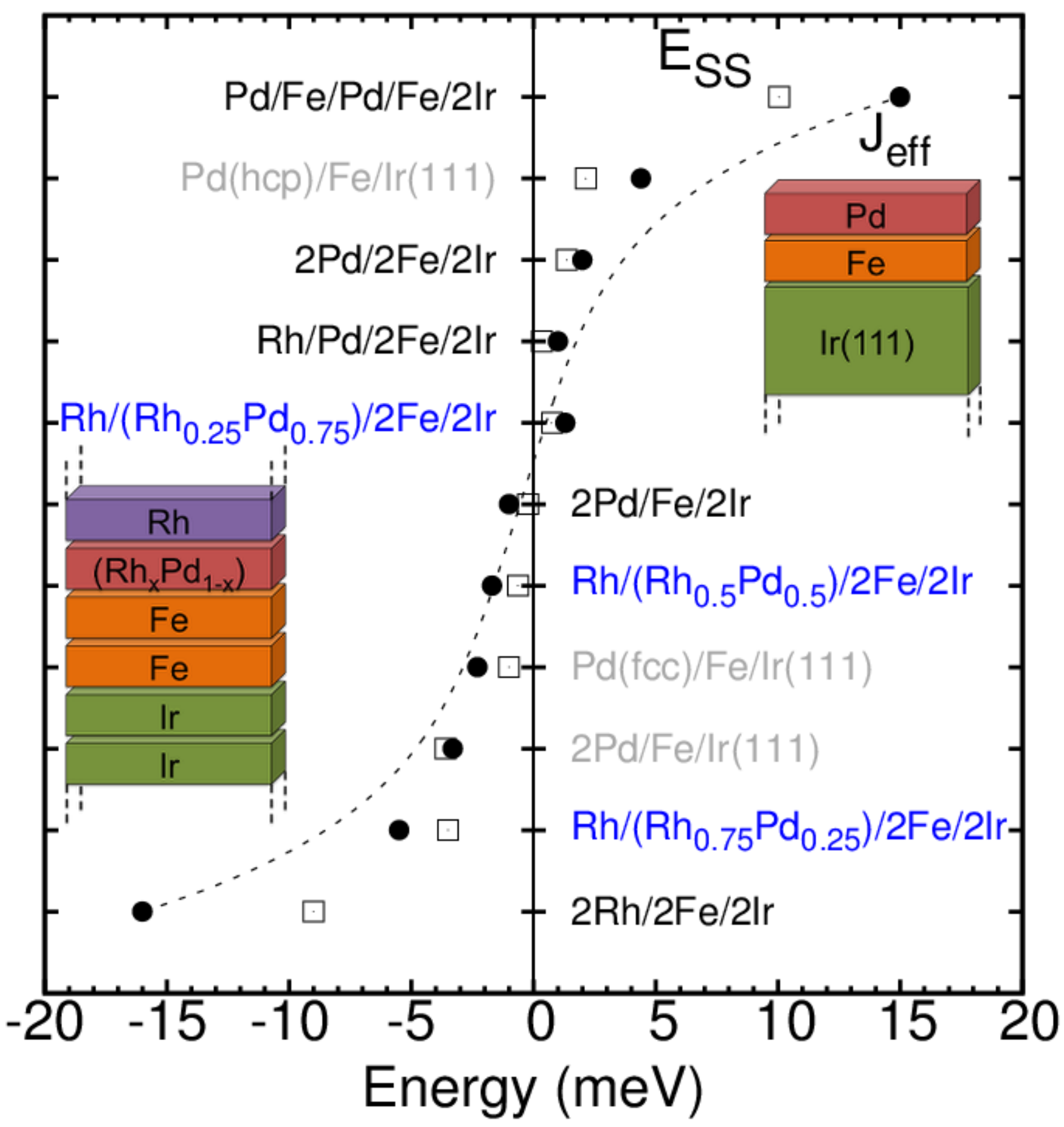}
\caption{Effective exchange interaction $J_{\rm eff}$ (full circles) in transition-metal multilayers 
(black/blue fonts) and ultra-thin films on surfaces (gray fonts). Insets show
sketches of the multilayer and thin film geometry. The energy difference between a 
spin spiral with a wave length of 2.7~nm with respect to the ferromagnetic state, $E_{\rm SS}$ 
is given by empty squares. The dashed line is a guide for the eye.}
\label{fig:Jeff}
\end{figure}

$J_{\rm eff}$ and $E_{\rm SS}$ are displayed for various systems in Fig.~\ref{fig:Jeff}. 
We start with the monolayer films on substrates which can host single skyrmions as shown 
in experiment~\cite{Romming:13.1} and explained by theory~\cite{Dupe:14.1}. For both fcc 
and hcp stacking of the Pd overlayer on Fe/Ir(111), we find a small value of the effective
exchange constant, i.e.~$J_{\rm eff}=-2.3$ and $+4.4$~meV, respectively.
The DMI which is of similar strength in both systems~\cite{Dupe:14.1} can stabilize
skyrmions in these films in an applied magnetic field. For two layers of Pd on Fe/Ir(111), $J_{\rm eff}$ is negative
and slightly stronger than for Pd(fcc)/Fe/Ir(111) and a skyrmion phase also occurs in
a magnetic field as found in our simulations. In these two systems, the spin spiral ground
state is already enforced by the exchange interaction and becomes even more favorable due
to the DMI. We can also conclude the spin spiral ground state from the negative
value of $E_{\rm SS}$ which closely follows the trend of $J_{\rm eff}$.
  
Now we turn to the multilayers which are built by an epitaxial fcc stacking of (111) layers with
a geometry as in an Ir bulk crystal but allowing for structural relaxation in the direction
normal to the layers (see methods).
We first consider a multilayer which is a repetition of a sandwich consisting of one atomic Fe layer 
between two Pd and two Ir layers denoted as 2Pd/Fe/2Ir. As shown in Fig.~\ref{fig:Jeff}, the effective
exchange interaction is very similar to that of the ultra-thin film system Pd(fcc)/Fe/Ir(111)
which demonstrates that $J_{\rm eff}$ is mainly controlled by the interfaces. A first na\"ive 
solution to increase the amount of magnetic material, i.e.~Fe layers in the sandwich, would be 
to repeat the Pd/Fe interface sequence as in Pd/Fe/Pd/Fe/2Ir. However, $J_{\rm eff}$ becomes much 
larger ($\approx 15$~meV) indicating strong ferromagnetic exchange interaction which prevents 
skyrmion formation (see supplement S4). If on the other
hand, we consider the multilayer with a basic unit consisting of 2 layers of Pd, Fe, and Ir,
i.e.~2Pd/2Fe/2Ir, the effective exchange interaction is of the same order as in the original
single Fe-layer system. 

We can understand the role of Pd by replacing it by Rh which has one electron less
in the $4d$ shell. For a multilayer consisting of 2Rh/2Fe/2Ir we obtain a value of 
$J_{\rm eff}\approx -16$~meV, i.e.~negative and of large absolute value. In this system, we find an exchange driven
spin spiral ground state and no skyrmion formation in a reasonable external magnetic field. By choosing
an intermediate interface composition, e.g.~Rh/Rh$_{0.75}$Pd$_{0.25}$/2Fe/2Ir, we can increase
the $4d$ band filling quasi continuously and thereby increase the value of the effective
exchange coupling $J_{\rm eff}$. For a large range of interface alloy compositions 
Rh$_x$Pd$_{1-x}$, we find multilayers with a small $J_{\rm eff}$ which can host skyrmionic ground states
as we show below using spin dynamics simulations.

After having revealed the feasibility to tune the magnetic interactions in transition-metal
multilayers, we demonstrate the formation of skyrmions in such systems.
In order to explore the magnetic phase space in an external magnetic field, we employ
the spin Hamiltonian
\begin{eqnarray}
H = & - & \sum_{ij} J_{ij} (\mathbf M_i \cdot \mathbf M_j)
      -   \sum_{ij} \mathbf D_{ij} \cdot (\mathbf M_i \times \mathbf M_j) \nonumber  \\
    & - & \sum_i K (M_i^z)^2
      -   \sum_i \mathbf B \cdot \mathbf M_i
\label{eq:modelH}
\end{eqnarray}
which describes the magnetic interactions between the magnetic moments $\mathbf{M}_i$ of atoms at sites $\mathbf{R}_i$.
The parameters for the exchange interaction ($J_{ij}$), the DMI ($\mathbf{D}_{ij}$)  as well as an uniaxial
magnetocrystalline anisotropy ($K$) were obtained from DFT.
 
In a magnetic bilayer 
it is convenient to split the exchange interaction into an intralayer contribution, $J_{\delta}^{\|}$, which 
corresponds to the exchange coupling to an atom of the $\delta$-th nearest neighbor within an Fe layer, 
and the interlayer part $J_{\delta}^{\perp}$, which corresponds 
to the coupling between the two Fe layers (see supplementary material S2 for details). 
For the multilayer 
Rh/Pd/2Fe/2Ir, on which we focus below as a generic example, the interlayer coupling between nearest neighbors
is very strong $J_{1}^{\perp}=24.7$~meV and favors ferromagnetic alignment of magnetic moments of
the two Fe layers.
However, the nearest-neighbor exchange coupling within each Fe layer, $J_{1}^{\|}=3.9$~meV,
is much smaller and on the order of antiferromagnetic exchange with third nearest neighbors, 
$J_{3}^{\|}=-3.0$~meV. 
In total, the competition between intra- and interlayer exchange leads to the small effective
exchange constant $J_{\rm eff}=1$~meV (see Fig.~\ref{fig:Jeff}), which indicates a very 
flat dispersion curve close to the FM state (cf.\ supplement S3). 

The DMI is of primary importance to stabilize skyrmions and has to be 
carefully evaluated for the bilayer case. Although there are contributions from the two 
interfaces, we find from our DFT calculations that the DMI is controlled by the 
Fe interface to the heavy $5d$ material (see supplement S3).
Within our spin Hamiltonian, we can describe it by an effective nearest-neighbor 
$\mathbf D$-vector acting on the Fe moments at the Fe/Ir interface only.
We obtain a value of $D=1.3$~meV for Rh/Pd/2Fe/2Ir with very little variation upon changing the composition of the
Rh$_x$Pd$_{1-x}$ interface (on the order of 0.1~meV). This value 
is very similar to those obtained for the ultra-thin film system Pd/Fe/Ir(111), i.e.~$D=1.0$~meV and $1.35$~meV 
for fcc and hcp stacking of Pd, respectively~\cite{Dupe:14.1}.
The strength of the DMI is sufficient to create a spin spiral ground state for Rh/Pd/2Fe/2Ir
with a right hand rotation sense and a period of $\lambda=2.25$~nm (cf.\ supplement S3).
The obtained uniaxial anisotropy, $K$, is $0.6$~meV per Fe atom and favors an out-of-plane magnetization.

\begin{figure}[thp]
\centering
\includegraphics[width=0.46\textwidth]{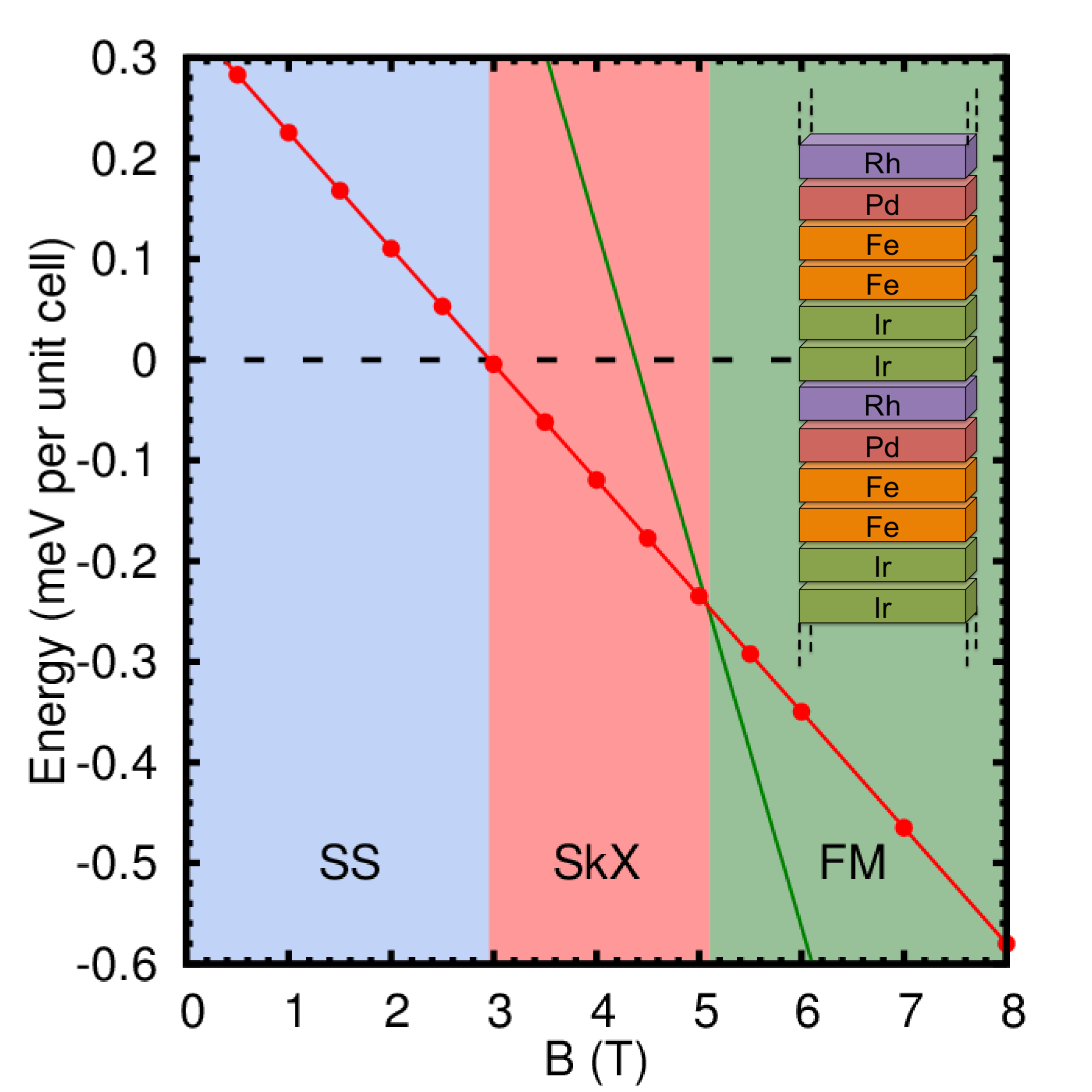}
\caption{Low temperature magnetic phase diagram of the Rh/Pd/2Fe/2Ir multilayer shown in the inset. The
total energy of the ferromagnetic (FM) state (green line) and of the skyrmion lattice (SkX, red line) is
shown with respect to the spin spiral (SS) state (dashed black line). The regimes of the SS, SkX, and FM
phase are indicated by blue, red, and green color, respectively. 
} \label{fig:diag-phase}
\end{figure}

We show the formation of skyrmions in multilayers by studying our spin Hamiltonian parametrized from first-principles 
using spin dynamics simulations (see methods). In
Fig.~\ref{fig:diag-phase}, the obtained low temperature phase diagram is shown as a function of an external magnetic 
field applied perpendicular to the layers.
At $B=0$~T the lowest energy is obtained for a spin spiral with a period of $\lambda=2.25$~nm. 
As the magnetic field increases, the spin spiral is destabilized and at $B=2.95$~T a transition to a skyrmion lattice occurs.
Fields of this strength have been used e.g.\ in Ref.\onlinecite{Romming:13.1} to induce skyrmions in monolayers experimentally
and further fine-tuning of the materials in the multilayer can be used to get to even lower critical fields.
At $B=5.1$~T, the magnetic field becomes large enough to stabilize the FM state. 

Since the DMI does not vary much upon changing the composition $x$
of the Rh/Rh$_x$Pd$_{1-x}$/2Fe/2Ir multilayer and the influence of the magnetocrystalline anisotropy energy 
is rather moderate, the magnetic field strengths at the phase transitions
depend mainly on the effective exchange interaction $J_{\rm eff}$. Although this interaction
is influenced by factors such as alloying at the interfaces, it can be tuned as shown in Fig.~\ref{fig:Jeff}
allowing a control over the phase diagram. (Limiting scenarios with respect to variation of the
exchange interaction and magnetocrystalline anisotropy are discussed in the supplement S4.)

\begin{figure}[thp]
\centering
\includegraphics[width=0.46\textwidth]{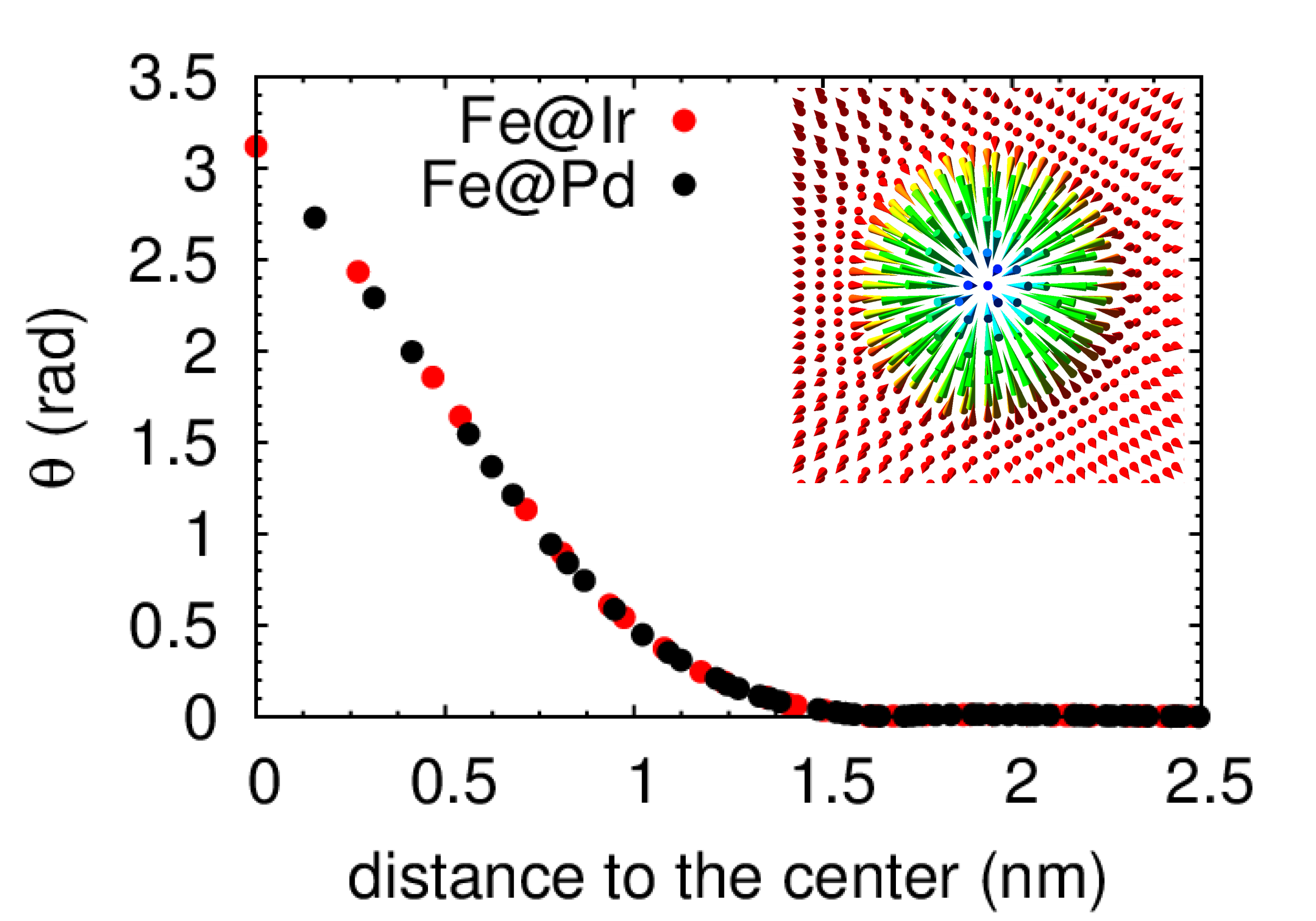}
\caption{Magnetization profile of an isolated skyrmion in the multilayer Rh/Pd/2Fe/2Ir averaged at a given 
radial distance $r$ from the center. The polar angle $\theta$ of the magnetization vector is shown
as a function of $r$. Red dots correspond to the Fe atoms at the Fe/Ir interface and the black dots 
correspond to the Fe atoms on the Pd side. 
The inset shows a top view of the magnetization of the skyrmion with red arrows pointing 
up and green/blue arrows pointing down.} \label{fig:profile}
\end{figure}

In Fig.~\ref{fig:profile} we present the profile of an isolated skyrmion in the bilayer at $B=3.5$~T, i.e.~the 
polar angle $\theta$ of the local magnetization as a function of the distance from the center. The curves for both 
Fe layers show that at the center of the skyrmion the magnetic moments are antiparallel to the surrounding FM background.
We obtain a skyrmion diameter of approximately 2.2~nm consistent with the spin spiral minimum.
The similarity between the two curves indicates that the magnetic moments in the 
two layers are strongly coupled and that the skyrmion exists in both layers simultaneously.

\begin{figure}[thp]
\centering
\includegraphics[width=0.46\textwidth]{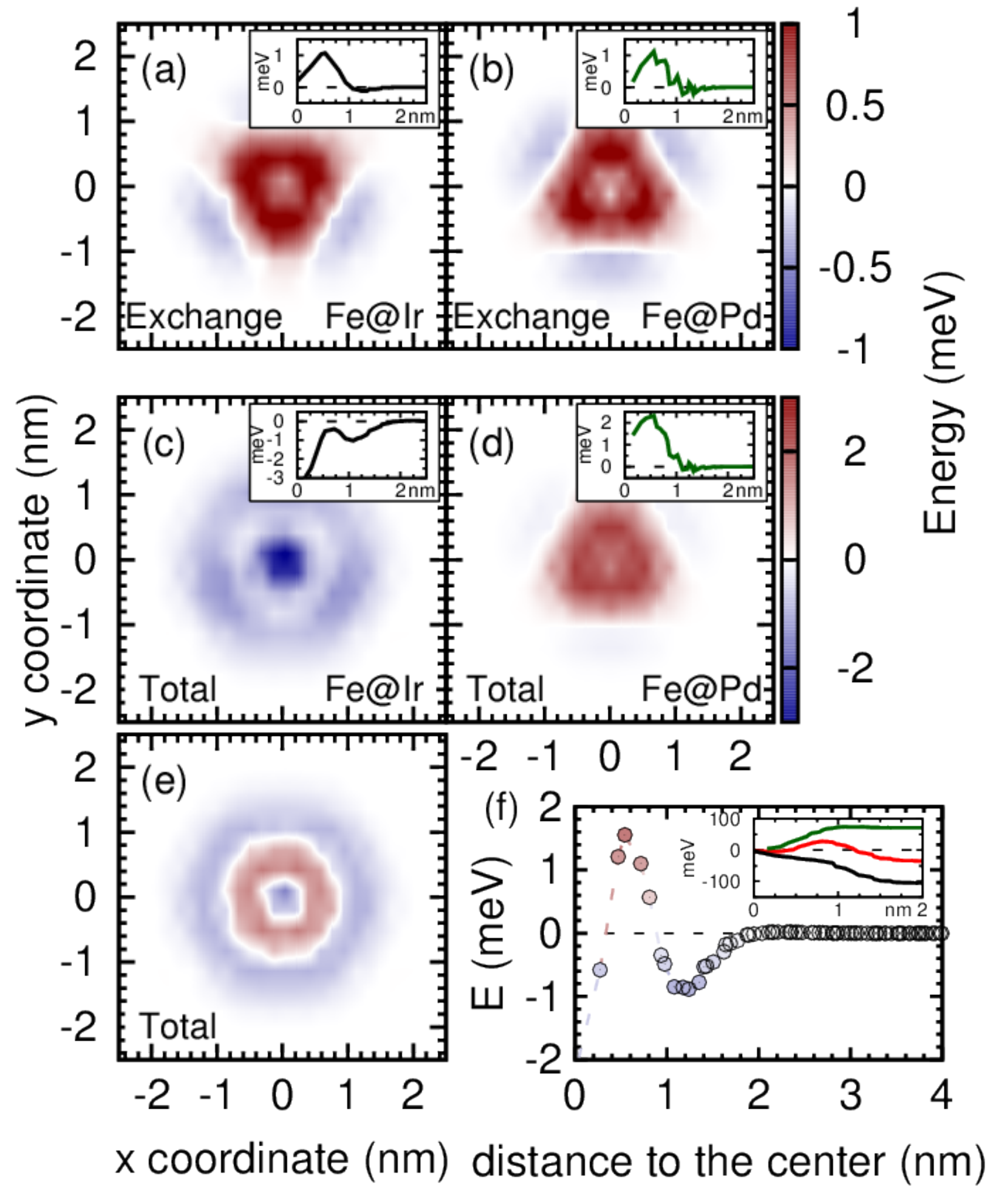}
\caption{Energy density contributions for an isolated skyrmion in an Rh/Pd/2Fe/2Ir multilayer at a 
magnetic field of $B=3.5$~T. (a,b) show the exchange energy in the Fe layer at the Ir interface (Fe@Ir)
and at the Pd interface (Fe@Pd), respectively. (c,d) show the total energy (exchange, DMI, Zeeman and 
anisotropy) for the two Fe layers and (e) for the complete Fe bilayer. The color scale bar of the 
energy is given to the right of the panels indicating stabilizing ($E < 0$) and destabilizing ($E > 0$)
contributions. For (e) the scale bar is the same as for (c) and (d).
The insets show an average of the energy density over the $\delta^{\rm th}$ neighbor shell. 
(f) total energy density averaged over the $\delta^{\rm th}$ neighbor shell for (e). The inset shows 
the integrated total energy of the Fe bilayer (red), Fe@Pd (green) and Fe@Ir (black).
} \label{fig:contributions}
\end{figure}

In order to understand the stabilization mechanism of a skyrmion in the Fe bilayer, we study 
the contributions of the different interactions to the energy density with respect to the 
ferromagnetic background in each Fe layer for the isolated skyrmion at $B=3.5$~T. 
The energy density due to exchange (Figs.~\ref{fig:contributions}a,b) is very similar for the two Fe layers
apart from a rotation by $60^{\circ}$ due to the different position of the first nearest neighbor atoms 
in the adjacent layer which are coupled by $J_1^{\perp}$ giving the main contribution to the exchange.
The exchange energy density is minimal in the skyrmion center where a moderate rotation of the magnetization (cf.~Fig.~\ref{fig:profile})
corresponding to a long wavelength spin spiral is energetically favored. 
As we move away from the center of the skyrmion, the canting between nearest-neighbor spins and
the exchange energy increases, reaches a maximum at a distance of $1$~nm from the core and decreases 
again when reaching the FM background. 

The total energy density reveals the difference between the two Fe layers.
For the Fe layer at the Pd interface (Fig.~\ref{fig:contributions}d) the DM
interaction vanishes, and the total energy density is only increased by the Zeeman term 
and the anisotropy contribution. 
However, the Fe layer at the Ir interface (Fig.~\ref{fig:contributions}c) is subject to a significant DMI and 
its energy contribution outweighs the anisotropy and the Zeeman term leading to a stabilizing (negative) 
energy density within the entire skyrmion.  The total energy density of the Fe bilayer, i.e.~the sum of 
the two layers, shown in Figs.~\ref{fig:contributions}e and f, exhibits a small local minimum at the center of the 
skyrmion due to the compensation of the DMI from the Fe/Ir interface and the Zeeman and anisotropy terms. 
As we move away from the center, the total energy density first increases due to the strong exchange and
Zeeman contributions and the rise of the anisotropy term and decreases beyond approximately 1~nm due to
the DM term (cf.~Fig.~\ref{fig:contributions}f).

The energy density integrated over a circle centered at the skyrmion is shown as a function of the radius $r$
for the Fe bilayer and the two Fe layers in the inset of Fig.~\ref{fig:contributions}f. For the Fe layer at 
the Pd interface (green curve), the integrated energy is always positive as expected from the energy profile 
(inset of Fig.~\ref{fig:contributions}d) and saturates at approximately $r=1.25$~nm. For the Fe layer at the
Ir interface (black curve), the energy is negative indicating a skyrmion stabilization due to the DMI.
The competition of these two contributions leads to the profile for the Fe bilayer (red curve) which displays a 
positive maximum at small radius but converges to a negative value of about $-32$~meV indicating that the skyrmion
is energetically favorable due to the DM contribution at the rim of the skyrmion.

We have focused on the system composed of an Fe bilayer sandwiched between Rh/Pd and Ir layers. 
For Fe monolayers on $4d$- and $5d$-transition metal surfaces we have demonstrated before that the 
exchange interaction can be tuned by the substrate band filling
from ferro- to antiferromagnetic \cite{VCA3,Hardrat09.1,Al-Zubi:11.2}. Therefore, other TM multilayers,
e.g.~Ru$_{x}$Pd$_{1-x}$/2Fe/Pt allow a similar tuning of $J_{\rm eff}$
by varying the composition $x$. The strength of the DMI, on the other hand, will be determined by the 
interface with the $5d$-TM, e.g.\ Ir or Pt~\cite{Wu,Hrabec2014}. 
Transition-metal multilayers consisting of a repetition of a few atomic layers 
as proposed here have recently been realized experimentally to stabilize chiral domain walls~\cite{Chen2015}. 

In conclusion, we have shown that magnetic skyrmions can emerge in transition-metal multilayers
consisting of Fe bilayers sandwiched between $4d$- and $5d$-transition metals. 
Such systems provide a rich field for the formation of skyrmions
and allow to go beyond the monolayer systems discussed so far \cite{Heinze:11.1,Romming:13.1,Dupe:14.1} 
to a thickness range where the unique transport properties of these systems can be studied and exploited 
in devices. 
Recent success in tailoring the DMI at the interface to the $5d$ material~\cite{Fert2015} can be combined
with our concept to tune the exchange interaction by the choice of the $4d$ component of the multilayers,
thus opening  a route to engineer skyrmion properties  for spintronic applications.

\section*{Methods}

We have explored the multilayer structures from first-principles based on the
full-potential linearized augmented plane wave method as implemented in the
FLEUR code~\cite{FLEUR}. Within this approach we can calculate the total energy 
of non-collinear magnetic structures such as spin spirals~\cite{Kurz-SS} including 
the DMI in first order perturbation theory with respect to the spin-orbit coupling~\cite{Heide-DMI}. 
We have used a two-dimensional hexagonal p$(1\times1)$ unit cell within each layer and 
the in plane-lattice parameter of the Ir(111) surface as obtained from DFT in
Ref.~[\onlinecite{Dupe:14.1}]. The distances between the different layers were relaxed 
using the mixed functional 
suggested in Ref.~[\onlinecite{mixed-fun}], which is ideally suited for interfaces 
of $3d$- and $5d$-transition metals. 
The magnetic properties were obtained within the local density approximation~\cite{doi-10.1139/p80-159}. 
In order to study alloying within the RhPd layers, we have used the virtual 
crystal approximation (VCA)~\cite{VCA2} which has been used successfully for various 
types of systems from solid solutions~\cite{VCA1} to magnetic ultrathin films~\cite{VCA3}.
Further computational details can be found in the supplementary material (S1).

In order to calculate the different energy differences between the FM, SkX and SS phases, we have relaxed a spin lattice
of the bilayer with (100$\times$100) spins in each layer according to the Landau-Lifshitz equation of spin dynamics:

\begin{equation}
\hbar \frac{\mathbf M_i}{dt} = \frac{\partial H_i}{\partial \mathbf M_i} \times \mathbf M_i - \beta \left( \frac{\partial H_i}{\partial \mathbf M_i} \times \mathbf M_i \right)  \times \mathbf M_i
\end{equation}

where $\beta$ is the damping term and $H_i$ is the Hamiltonian of spin $i$ and is expressed by the Hamiltonian given
in Eq.~(1) if we omit the sum over $i$. In order to achieve fast relaxation, we have used $\beta=0.95$. We have carried 
out simulations on a micro-second time scale with time steps ranging from 0.1 ps to 10 ps . The equation of motion was 
integrated with a simple Euler, a Heun, and two semi-implicit integrators~\cite{Mentink2010} leading to the same results. 
To ensure that we reached the ground state, we have simultaneously run Monte Carlo simulations with a Metropolis algorithm 
on these configurations at $T=0$~K. Depending on the system, both techniques gives similar results with a precision on the 
order of a few $\mu{\rm eV/spin}$, sufficient to discriminate between the different phases in our cases.

\section*{Acknowledgments}
Research leading to these results has received funding from the Deutsche Forschungsgemeinschaft 
via project DU1489/2-1. We gratefully acknowledge computing time at the JUROPA supercomputer 
from the J\"{u}lich Supercomputing Centre and the supercomputer of the
North-German Supercomputing Alliance (HLRN).

\bibliographystyle{apsrev4-1}

\bibliography{text}

\end{document}